\documentclass[aps,prd,onecolumn,showpacs,showkeys,amsmath,amssymb]{revtex4}
\usepackage{amsfonts}
\usepackage{graphicx}
\usepackage{dcolumn}
\usepackage{bm}
\usepackage[dvips]{color}
\makeatletter
\newcommand{\figcaption}{\def\@captype{figure}\caption}
\newcommand{\tabcaption}{\def\@captype{table}\caption}
\makeatother

\begin{document}


\title{Possible $J^{PC} = 0^{--}$ Charmonium-like State}

\author{Wei Chen}
\email{boya@pku.edu.cn} \affiliation{Department of Physics
and State Key Laboratory of Nuclear Physics and Technology\\
Peking University, Beijing 100871, China  }
\author{Shi-Lin Zhu}
\email{zhusl@pku.edu.cn} \affiliation{Department of Physics
and State Key Laboratory of Nuclear Physics and Technology\\
and Center of High Energy Physics, Peking University, Beijing
100871, China }

\begin{abstract}

We study the possible charmonium-like states with $J^{PC}=0^{--},
0^{-+}$ using the tetraquark interpolating currents with the QCD sum
rules approach. The extracted masses are around $4.5$ GeV for the
$0^{--}$ charmonium-like state and $4.6$ GeV for the $0^{-+}$
charmonium-like state while their bottomonium-like analogues lie
around $10.6$ GeV. We also discuss the possible decay, production,
and experimental search of the $0^{--}$ charmonium-like state.

\end{abstract}



\maketitle

\pagenumbering{arabic}

\section{Introduction}\label{sec:intro}
%
Since the Belle Collaboration observed the narrow state $X(3872)$ on
the threshold of $D^0\bar{D^0}^{\ast}$ in the $B^+\rightarrow
K^+X(3872)\rightarrow K^+J/\psi\pi^+\pi^-$ channel in 2003
\cite{2003-Choi-p262001-262001}, many new charmonium or
charmonium-like states have been observed such as $Y(3940), Y(4260),
Z(3930), X(3940), Y(4325), Y(4360), Y(4660), Z^+(4430), Z^+(4050),
Z^+(4250)$ and $Y(4140)$ \cite{2005-Choi-p182002-182002,
2005-Aubert-p142001-142001, 2006-Uehara-p82003-82003,
2007-Abe-p82001-82001, 2007-Aubert-p212001-212001,
2007-Yuan-p182004-182004, 2007-Wang-p142002-142002,
2008-Choi-p142001-142001, 2008-Mizuk-p72004-72004,
2009-Aaltonen-p242002-242002}. For experimental reviews, one can
consult Refs. \cite{2006-Swanson-p243-305, 2008-Zhu-p283-322,
2008-Amsler-p1, 2009-Bracko-p-}.

The discovery of these new states have enriched the charmonium
spectroscopy greatly. It is very difficult to accommodate all these
states in the conventional quark model. In order to study their
underlying structure, many interpretations were proposed such as the
hybrid mesons, the molecular or tetraquark states, baryonium states
and so on. For example, $X(3872)$ was speculated to be a hybrid
charmonium state in Ref. \cite{2003-Close-p210-216}, a
$D\bar{D}^{\ast}$ molecular state in
Ref.\cite{2004-Swanson-p197-202} and a $cq\bar c\bar q$ tetraquark
state in Ref. \cite{2007-Matheus-p14005-14005}. $Z^+(4430)$ was
assigned as a $D^{\ast}\bar{D_1}$ molecular state in Ref.
\cite{2008-Liu-p94015-94015} and a $cs\bar c\bar s$ tetraquark state
in Ref. \cite{2009-Bracco-p240-244}. $Y(4260)$ was proposed as a
hybrid charmonium state in Ref. \cite{2005-Close-p215-222} and a
$cq\bar c\bar q$ tetraquark state in Ref.
\cite{2005-Maiani-p031502-031502}. $Y(4140)$ was proposed as a
$D^{\ast}_s\bar{D^{\ast}_s}$ molecular state in Ref.
\cite{2009-Liu-p17502-17502}. However, one should be very cautious
that the conventional charmonium spectrum may be distorted if one
considers either the coupled-channel effect or the screened linear
confinement force \cite{libaiqing}. More charmonium states can be
accommodated below 5 GeV within this picture \cite{libaiqing}.

The possible $c\bar cq\bar q$ states with various quantum numbers
including $J^{PC}=0^{--}$ were investigated extensively in Ref.
\cite{1980-Chao-p281-281} in the early 1980s. The author discussed
the spectroscopy, decay and production of the $c\bar cq\bar q$
systems with the angular momentum $L\geq1$ by taking account of
the color magnetic and electric forces. In the conventional quark
model, states with $J^{PC}=0^{--}$ are exotic states. They cannot
be composed of a pair of quark and antiquark. In Ref.
\cite{2009-Jiao-p114034-114034}, we noticed that the light
tetraquark currents with $J^{PC}=0^{--}$ do not support a
low-lying resonant signal below 2 GeV. Since increasing the quark
mass reduces the kinetic energy and thus may help stabilize the
system, we will study whether the $c\bar cq\bar q$ states with
$J^{PC}=0^{--}$ exist in this paper.

The paper is organized as follows. We construct the tetraquark
currents with $J^{PC}=0^{--}, 0^{-+}$ using the diquark and
antidiquark fields in Sec. II and derive the spectral densities in
Sec. III which are collected in the Appendix. We perform the
numerical analysis and extract the masses in Sec. IV and discuss the
possible decay, production, and experimental search of the $0^{--}$
charmonium-like states in the last section.

%
%
\section{TETRAQUARK INTERPOLATING CURRENTS}\label{currents}

We have constructed the light tetraquark interpolating currents
with $J^{PC}=0^{--}$ using the diquark-antidiquark fields in the
previous work \cite{2009-Jiao-p114034-114034}. In this work we
follow the same steps. We first construct ten color singlet
pseudoscalar operators considering both the Lorentz and color
structures: $S_6, P_6, V_6, A_6, T_6, S_3, P_3, V_3, A_3, T_3$.
The subscripts $\mathbf 6$ and $\mathbf 3$ indicate that the color
structures of the tetraquark are $\mathbf {6\otimes\bar6}$ and
$\mathbf {\bar3\otimes3}$ respectively. Details can be found in
Ref.\cite{2009-Jiao-p114034-114034}. With the charge-conjugation
transformation we get
\begin{equation}
\mathbb{C}S_6\mathbb{C}^{-1}=V_6 \, ,
\mathbb{C}A_6\mathbb{C}^{-1}=P_6 \, ,
\mathbb{C}A_3\mathbb{C}^{-1}=P_3 \, ,
\mathbb{C}S_3\mathbb{C}^{-1}=V_3 \,
,\mathbb{C}T_6\mathbb{C}^{-1}=T_6 \, ,
\mathbb{C}T_3\mathbb{C}^{-1}=T_3 \, .
\end{equation}
Using the above charge-conjugation relations, we can obtain the
currents with definite $C$ parity:
\begin{enumerate}
\item For the quantum number $J^{PC}=0^{--}$:
\begin{eqnarray}
\nonumber \eta_1 &=&
S_6-V_6=q_{a}^TCc_{b}(\bar{q}_{a}\gamma_5C\bar{c}^T_{b}+\bar{q}_{b}\gamma_5C\bar{c}^T_{a})
-q_{a}^TC\gamma_5c_{b}(\bar{q}_{a}C\bar{c}^T_{b}+\bar{q}_{b}C\bar{c}^T_{a})
\, ,
\\ \nonumber
\eta_2 &=&
A_6-P_6=q_{a}^TC\gamma_{\mu}c_{b}(\bar{q}_{a}\gamma^{\mu}\gamma_5C\bar{c}^T_{b}+\bar{q}_{b}\gamma^{\mu}
\gamma_5C\bar{c}^T_{a})-q_{a}^TC\gamma_{\mu}\gamma_5c_{b}(\bar{q}_{a}\gamma^{\mu}C\bar{c}^T_{b}+\bar{q}_{b}
\gamma^{\mu}C\bar{c}^T_{a}) \, ,
\\   \label{currents1}
\eta_3 &=&
A_3-P_3=q_{a}^TC\gamma_{\mu}c_{b}(\bar{q}_{a}\gamma^{\mu}\gamma_5C\bar{c}^T_{b}-\bar{q}_{b}\gamma^{\mu}
\gamma_5C\bar{c}^T_{a})-q_{a}^TC\gamma_{\mu}\gamma_5c_{b}(\bar{q}_{a}\gamma^{\mu}C\bar{c}^T_{b}-\bar{q}_{b}
\gamma^{\mu}C\bar{c}^T_{a}) \, ,
\\ \nonumber
\eta_4 &=&
S_3-V_3=q_{a}^TCc_{b}(\bar{q}_{a}\gamma_5C\bar{c}^T_{b}-\bar{q}_{b}\gamma_5C\bar{c}^T_{a})
-q_{a}^TC\gamma_5c_{b}(\bar{q}_{a}C\bar{c}^T_{b}-\bar{q}_{b}C\bar{c}^T_{a})
\, .
\end{eqnarray}
\item For the quantum number $J^{PC}=0^{-+}$:
\begin{eqnarray}
\nonumber \eta_5 &=&
S_6+V_6=q_{a}^TCc_{b}(\bar{q}_{a}\gamma_5C\bar{c}^T_{b}+\bar{q}_{b}\gamma_5C\bar{c}^T_{a})
+q_{a}^TC\gamma_5c_{b}(\bar{q}_{a}C\bar{c}^T_{b}+\bar{q}_{b}C\bar{c}^T_{a})
\, ,
\\ \nonumber
\eta_6 &=&
T_3=q_{a}^TC\sigma_{\mu\nu}c_{b}(\bar{q}_{a}\sigma^{\mu\nu}\gamma_5C\bar{c}^T_{b}-\bar{q}_{b}\sigma^{\mu\nu}
\gamma_5C\bar{c}^T_{a}) \, ,
\\ \nonumber
\eta_7 &=&
A_6+P_6=q_{a}^TC\gamma_{\mu}c_{b}(\bar{q}_{a}\gamma^{\mu}\gamma_5C\bar{c}^T_{b}+\bar{q}_{b}\gamma^{\mu}
\gamma_5C\bar{c}^T_{a})+q_{a}^TC\gamma_{\mu}\gamma_5c_{b}(\bar{q}_{a}\gamma^{\mu}C\bar{c}^T_{b}+\bar{q}_{b}
\gamma^{\mu}C\bar{c}^T_{a}) \, ,
\\ \label{currents2}
\eta_8 &=&
A_3+P_3=q_{a}^TC\gamma_{\mu}c_{b}(\bar{q}_{a}\gamma^{\mu}\gamma_5C\bar{c}^T_{b}-\bar{q}_{b}\gamma^{\mu}
\gamma_5C\bar{c}^T_{a})+q_{a}^TC\gamma_{\mu}\gamma_5c_{b}(\bar{q}_{a}\gamma^{\mu}C\bar{c}^T_{b}-\bar{q}_{b}
\gamma^{\mu}C\bar{c}^T_{a}) \, ,
\\ \nonumber
\eta_9 &=&
S_3+V_3=q_{a}^TCc_{b}(\bar{q}_{a}\gamma_5C\bar{c}^T_{b}-\bar{q}_{b}\gamma_5C\bar{c}^T_{a})
+q_{a}^TC\gamma_5c_{b}(\bar{q}_{a}C\bar{c}^T_{b}-\bar{q}_{b}C\bar{c}^T_{a})
\, ,
\\ \nonumber
\eta_{10} &=&
T_6=q_{a}^TC\sigma_{\mu\nu}c_{b}(\bar{q}_{a}\sigma^{\mu\nu}\gamma_5C\bar{c}^T_{b}+\bar{q}_{b}\sigma^{\mu\nu}
\gamma_5C\bar{c}^T_{a}) \, .
\end{eqnarray}
\end{enumerate}
It is understood that Eqs.~(\ref{currents1})-(\ref{currents2})
should contain $(uc\bar u \bar c + dc\bar d \bar c)$ in order to
have definite isospin and $G$-parity. Because of the SU(2) flavor
symmetry, we do not differentiate the $up$ and $down$ quarks in our
analysis and denote them by $q$.

%
%
\section{QCD SUM RULE}\label{sec:QSR}
QCD sum rule is a powerful approach to study the hadron properties
in the past several decades\cite{M.A.Shifman:1979wz,
L.J.Reinder:1985wz, Colangelo:2000dp}. We consider the two-point
correlation function:
\begin{equation}
\Pi(q^{2})\equiv\int d^4
xe^{iqx}\langle0|T\eta(x)\eta^\dag(0)|0\rangle, \label{equ:Pi}
\end{equation}
where $\eta$ is an interpolating current. At the hadron level, the
correlation function $\Pi(q^{2})$ is expressed via the dispersion
relation:
\begin{eqnarray}
\Pi^{phen}(p^2)=\int_0^{\infty}\frac{\rho^{phen}(s)}{s-p^2-i\epsilon}.
\label{Phenpi}
\end{eqnarray}
The spectral function reads:
\begin{eqnarray}
\nonumber
\rho^{phen}(s)&\equiv&\sum_n\delta(s-m_n^2)\langle0|\eta|n\rangle\langle n|\eta^+|0\rangle\\
&=&f_X^2\delta(s-m_X^2)+ \mbox{continuum},   \label{Phenrho}
\end{eqnarray}
where the usual pole plus continuum parametrization of the hadronic
spectral density is adopted. $f_X$ is the overlapping parameter of
the current to the pseudoscalar state $X$:
$\langle0|\eta|X\rangle=f_X.$

At the quark-gluon level, the spectral density can be evaluated up
to dimension eight with the same method in
Refs.\cite{2007-Matheus-p14005-14005, 2009-Bracco-p240-244,
Matheus:2009vq, Zhang:2009vs}. Omitting the light quark mass, we use
the coordinate-space expression for the light quark propagator and
momentum-space expression for the charm quark propagator:
\begin{eqnarray}
\nonumber iS^{ab}_q(x)=  \frac{i\delta^{ab}}{2\pi^2x^4}\hat{x}
+\frac{i}{32\pi^2}\frac{\lambda^n_{ab}}{2}g_sG_{\mu\nu}^n\frac{1}
{x^2}(\sigma^{\mu\nu}\hat{x}+\hat{x}\sigma^{\mu\nu})
-\frac{\delta^{ab}}{12}\langle\bar{q}q\rangle+
\frac{\delta^{ab}x^2}{192}\langle\bar qg_s\sigma\cdot Gq\rangle,
\end{eqnarray}
\begin{eqnarray}
iS^{ab}_c(p) &=&
\frac{i\delta^{ab}}{\hat{p}-m_c}+\frac{i}{4}g_s\frac{\lambda^n_{ab}}{2}G_{\mu\nu}^n
\frac{\sigma^{\mu\nu}(\hat{p}+m_c)+(\hat{p}+m_c)\sigma^{\mu\nu}}
{(p^2-m_c^2)^2}+\frac{i\delta^{ab}}{12}\langle g_s^2GG\rangle
m_c\frac{p^2+m_c\hat{p}}{(p^2-m_c^2)^4}
\\ \nonumber
&&-\frac{i\delta^{ab}}{48}\langle
g_s^3fGGG\rangle\frac{(p^2+7m_c^2)\hat{p}+2m_c(3p^2+m_c^2)}{(p^2-m_c^2)^5}
\end{eqnarray}
where $\hat{x}\equiv\gamma_{\mu}x^{\mu}$,
$\hat{p}\equiv\gamma_{\mu}p^{\mu}$, $\langle\bar qg_s\sigma\cdot
Gq\rangle= \langle g_s \bar{q}\sigma^{\mu\nu} G_{\mu\nu}q\rangle$,
$\langle g_s^2GG\rangle=\langle g_s^2G_{\mu\nu}G^{\mu\nu}\rangle$,
$\langle g_s^3fGGG\rangle=\langle
g_s^3f^{abc}G^a_{\gamma\delta}G^b_{\delta\epsilon}G^c_{\epsilon\gamma}\rangle$,
$a$ and $b$ are the color indices. The momentum-space propagator
with three soft gluon lines can be found in Ref.
\cite{L.J.Reinder:1985wz}. For the light quark propagator, we use
the $D$ dimension coordinate-space expression. The dimensional
regularization is used throughout our calculation. The $\Pi(q^{2})$
in the operator product expansion(OPE) side can be written as:
\begin{eqnarray}
\Pi^{OPE}(p^2)=\int_{4m_c^2}^{\infty}\frac{\rho^{OPE}(s)}{s-p^2-i\epsilon},
\label{OPEPI}
\end{eqnarray}
Performing Borel transformation for the correlation function, we
arrive at:
\begin{eqnarray}
f_X^2e^{-m_X^2/M_B^2}=\int_{4m_c^2}^{s_0}dse^{-s/M_B^2}\rho^{OPE}(s).
\label{sumrule}
\end{eqnarray}
where $s_0$ is the threshold parameter. The mass $M_X$ reads:
\begin{eqnarray}
m_X^2=\frac{\int_{4m_c^2}^{s_0}dse^{-s/M_B^2}s\rho^{OPE}(s)}{\int_{4m_c^2}^{s_0}dse^{-s/M_B^2}\rho^{OPE}(s)}.
\label{mass}
\end{eqnarray}

\begin{figure}[hbtp]
\begin{center}
\begin{tabular}{lr}
\scalebox{0.6}{\includegraphics{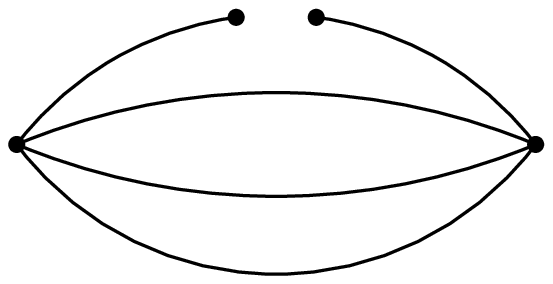}}&\hspace{1.5cm}
\scalebox{0.6}{\includegraphics{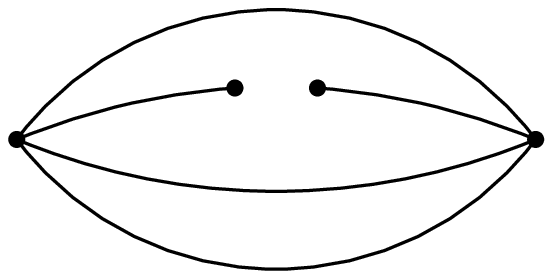}}\\
\scalebox{0.6}{\includegraphics{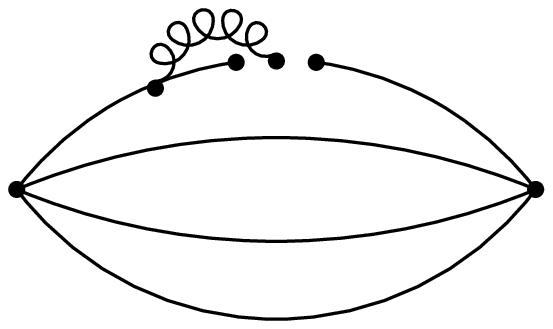}}&\hspace{1.5cm}
\scalebox{0.6}{\includegraphics{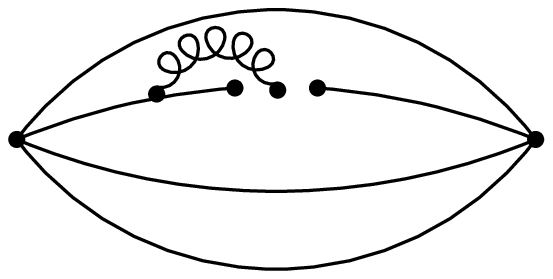}}\\
\scalebox{0.6}{\includegraphics{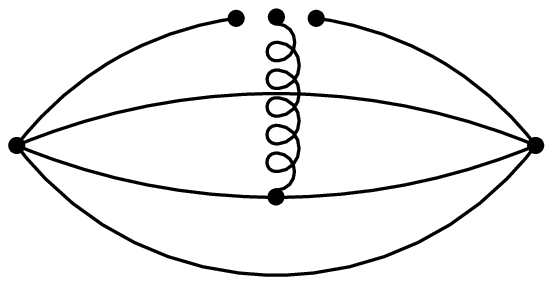}}&\hspace{1.5cm}
\scalebox{0.6}{\includegraphics{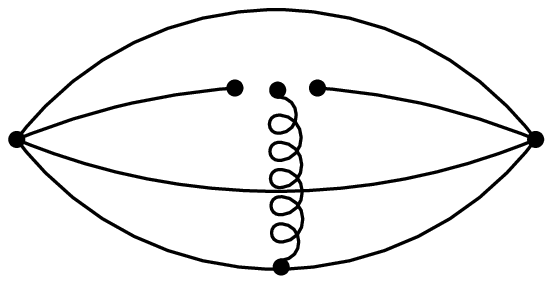}}
\end{tabular}
\caption{Feynman diagrams for $m_c\langle\bar qq\rangle$ and
$m_c\langle\bar qg_s\sigma\cdot Gq\rangle$. In each diagram, the two
upper lines represent the light quark propagators and the two lower
lines represent the charm quark propagators.} \label{fig1}
\end{center}
\end{figure}

For all the tetraquark currents in Eqs.~(\ref{currents1}) and
(\ref{currents2}), we collect the $\rho^{OPE}(s)$ in the Appendix.
Both the quark condensate $\langle\bar qq\rangle$ and quark gluon
mixed condensate $\langle\bar qg_s\sigma\cdot Gq\rangle$ vanish in
the chiral limit $m_q=0$. One may wonder whether the quark
condensates proportional to the charm quark mass: $m_c\langle\bar
qq\rangle$ and $m_c\langle\bar qg_s\sigma\cdot Gq\rangle$ exist.
They are usually very important corrections in the scalar, vector
and axial-vector channels \cite{2007-Matheus-p14005-14005,
Zhang:2009vs, 2009-Lee-p29-39, 2009-Wang-p375-382}. However, these
terms also vanish in the pseudoscalar channel. The diagrams in
Fig.\ref{fig1} vanish due to the special Lorenz structures of the
currents.
%

\section{Numerical Analysis}\label{sec:num}
%
We use the following values of the quark masses and various
condensates in the QCD sum rule analysis \cite{2008-Amsler-p1,
M.A.Shifman:1979wz, 2001-Eidemuller-p203-210,
1999-Jamin-p300-303}. The charm and bottom quark masses are the
running masses in the $\overline{MS}$ scheme:
\begin{eqnarray}
\nonumber
&&m_c(m_c)=(1.23\pm0.09)\text{ GeV} \, ,
\\ \nonumber
&&m_b(m_b)=(4.20\pm0.07)\text{ GeV} \, ,
\\ \nonumber
&&\langle\bar qq\rangle=-(0.23\pm0.03)^3\text{ GeV}^3 \, ,
\\ \nonumber
&&\langle\bar qg_s\sigma\cdot Gq\rangle=-M_0^2\langle\bar
qq\rangle\, ,
\\ \nonumber
&&M_0^2=(0.8\pm0.2)\text{ GeV}^2 \, ,
\\
&&\langle\bar ss\rangle/\langle\bar qq\rangle=0.8\pm0.2 \, ,
\\ \nonumber
&&\langle g_s^2GG\rangle=0.88\text{ GeV}^4 \, ,
\\ \nonumber
&&\langle g_s^3fGGG\rangle=0.045\text{ GeV}^6 \, . \label{parameters}
\end{eqnarray}

%
\begin{center}
\begin{tabular}{lr}
\scalebox{0.67}{\includegraphics{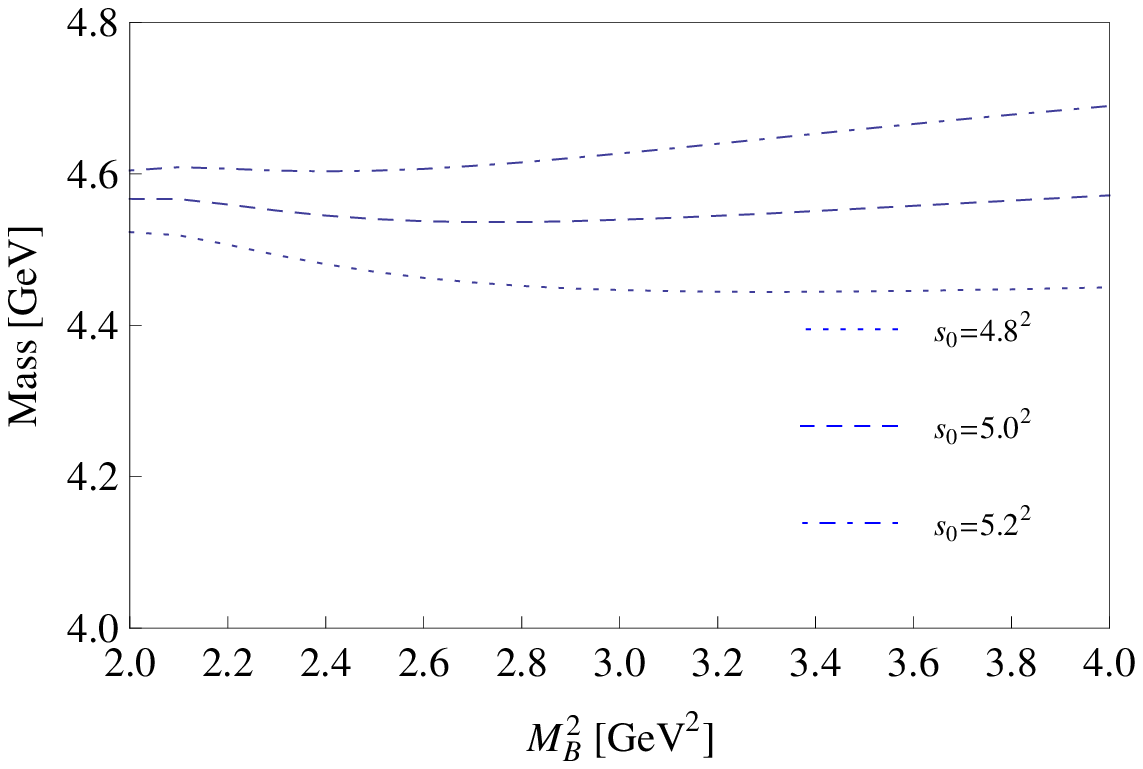}}&
\scalebox{0.67}{\includegraphics{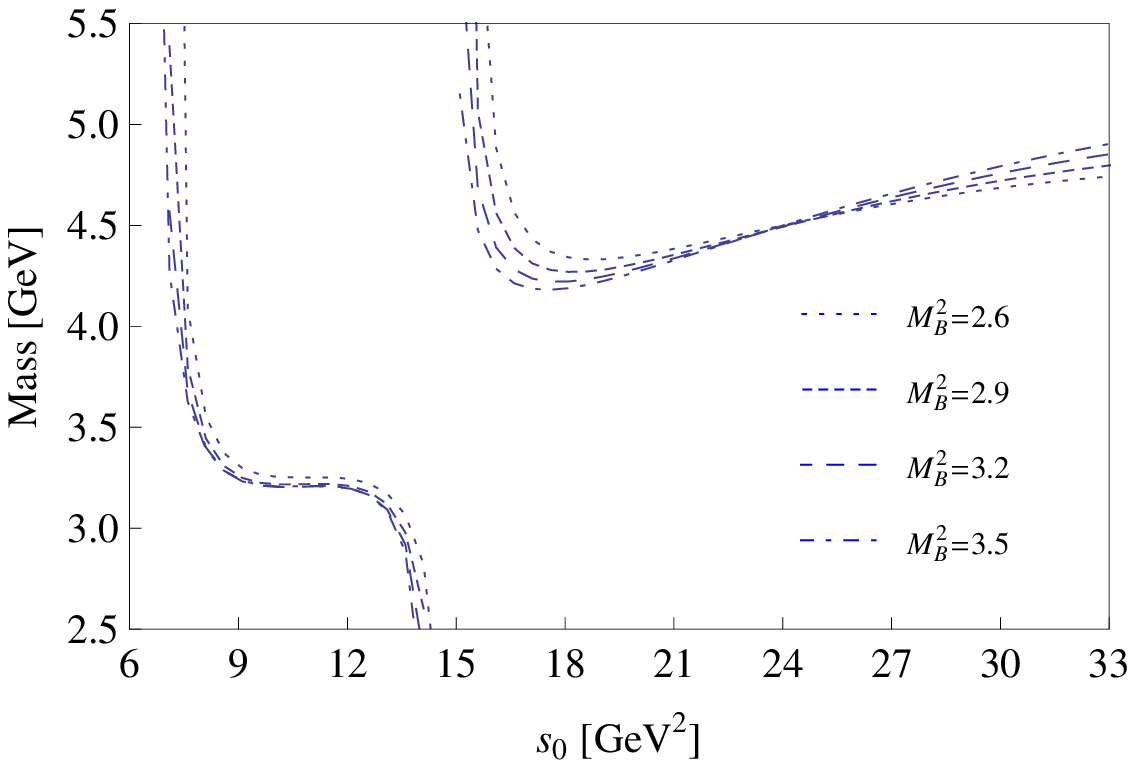}}
\end{tabular}
\figcaption{The variation of $M_X$ with $M^2_B$ (left) and $s_0$
(right) for the current $\eta_{2}$.} \label{mass2}
\end{center}
%

The Borel mass $M_B$ and the threshold value $s_0$ are two pivotal
parameters in the numerical analysis. The working region of the
Borel parameter is determined by the convergence of the operator
product expansion and the pole contribution. The requirement of the
convergence of OPE leads to the lower bound $M^2_{min}$ of the Borel
parameter. The upper bound $M^2_{max}$ of the Borel parameter
results from the requirement of the pole contribution.

The four quark condensate $\langle \bar qq\rangle$$^2$ is negative
and the most important condensate correction numerically. Its
absolute value is much bigger than the gluon condensate $\langle
g^2G^2\rangle$ and the dimension 8 condensate $\langle\bar
qg\sigma\cdot Gq\rangle\langle\bar qq\rangle$. Technically, we
require that the four quark condensate be less than one third of
the perturbative term to ensure the convergence of OPE, which
results at the lower bound of the Borel working window, $M^2_{min}
\sim 2.4$ GeV.

The pole contribution (PC) is defined as
\begin{eqnarray}
\frac{\int_{4m_c^2}^{s_0}dse^{-s/M_B^2}\rho(s)}{\int_{4m_c^2}^{\infty}dse^{-s/M_B^2}\rho(s)},
\label{pc}
\end{eqnarray}
which depends on both the Borel mass $M_B$ and the threshold value
$s_0$. $s_0$ is chosen around the region where the variation of
$m_X$ with $M_B^2$ is the minimum. For example, we choose $s_0\sim
25 $GeV$^2$ from the variation of the mass with $s_0$ as shown in
Figs. \ref{mass2} and \ref{mass6}. Requiring that the PC is larger
than $40\%$, one gets the upper bound $M_{max}^2$ of the Borel
parameter $M_B^2$. We list the working range of the Borel
parameter for the ten tetraquark currents $\eta_1\sim\eta_{10}$ in
table \ref{table1}. The masses are extracted using these threshold
values and $M_B^2=3.5$GeV$^2$, which are also listed in Table
\ref{table1}. Only the errors which arise from the uncertainty of
the threshold values and variation of the Borel parameter are
taken into account. Other possible error sources include the
truncation of the OPE series and the uncertainty of the condensate
values etc. The last column is the pole contribution.
%
\begin{center}
\begin{tabular}{c|c|c|c|c|c}
\hline \hline
                   & Currents & $s_0(\mbox{GeV}^2)$ & $[M^2_{\mbox{min}}$, $M^2_{\mbox{max}}](\mbox{GeV}^2)$& $m_X$\mbox{(GeV)}      & PC(\%)\\
\hline
                     & $\eta_1$      &  25                  & $2.4\sim3.6$                         & $-$              & - \\
                     & $\eta_2$      &  25                  & $2.4\sim3.7$                         & $4.55\pm0.11$    & 46.3  \\
$J^{PC}=0^{--}$      & $\eta_3$      &  25                  & $2.4\sim3.7$                         & $-$              & - \\
                     & $\eta_4$      &  25                  & $2.4\sim3.7$                         & $4.55\pm0.11$    & 45.9 \\
\hline
                     & $\eta_5$      &  25                  & $2.4\sim3.6$                         & $-$              & - \\
                     & $\eta_6$      &  27                  & $2.4\sim4.1$                         & $4.72\pm0.10$    & 53.8 \\
                     & $\eta_7$      &  25                  & $2.4\sim3.8$                         & $-$              & - \\
$J^{PC}=0^{-+}$      & $\eta_8$      &  25                  & $2.4\sim3.7$                         & $-$              & - \\
                     & $\eta_9$      &  25                  & $2.4\sim3.7$                         & $4.55\pm0.11$    & 45.9 \\
                     & $\eta_{10}$   &  27                  & $2.4\sim4.2$                         & $4.67\pm0.10$    & 56.8 \\
\hline \hline
\end{tabular}
\tabcaption{The threshold value, Borel window, mass and pole
contribution for $\eta_1\sim\eta_{10}$. The mass and pole
contribution are calculated at $M_B^2=3.5$GeV$^2$. \label{table1}}
\end{center}

From the variation of the mass with $s_0$, there is a plateau in
the region of $s_0=9$$\sim$$13$GeV$^2$ for $\eta_2, \eta_4,
 \eta_6, \eta_9, \eta_{10}$ as can be seen in Fig. \ref{mass2}.
This plateau looks like a resonance signal. However, it's just an
unphysical artifact because both the numerator and denominator in
Eq.(\ref{mass}) are negative within this region. The variation of
$M_X$ with the Borel parameter is weak.

For $\eta_1, \eta_3, \eta_5, \eta_7, \eta_8$, the extracted mass
$M_X$ grows monotonically with the threshold value. Also the
variation of $M_X$ with the Borel parameter is significant. So we
do not present the numerical values in Table \ref{table1} for
these currents. These currents may couple to the $0^{--}, 0^{-+}$
states very weakly. The continuum contribution may be quite large.
These two factors may lead to the above unstable mass sum rules.

\begin{center}
\begin{tabular}{lr}
\scalebox{0.66}{\includegraphics{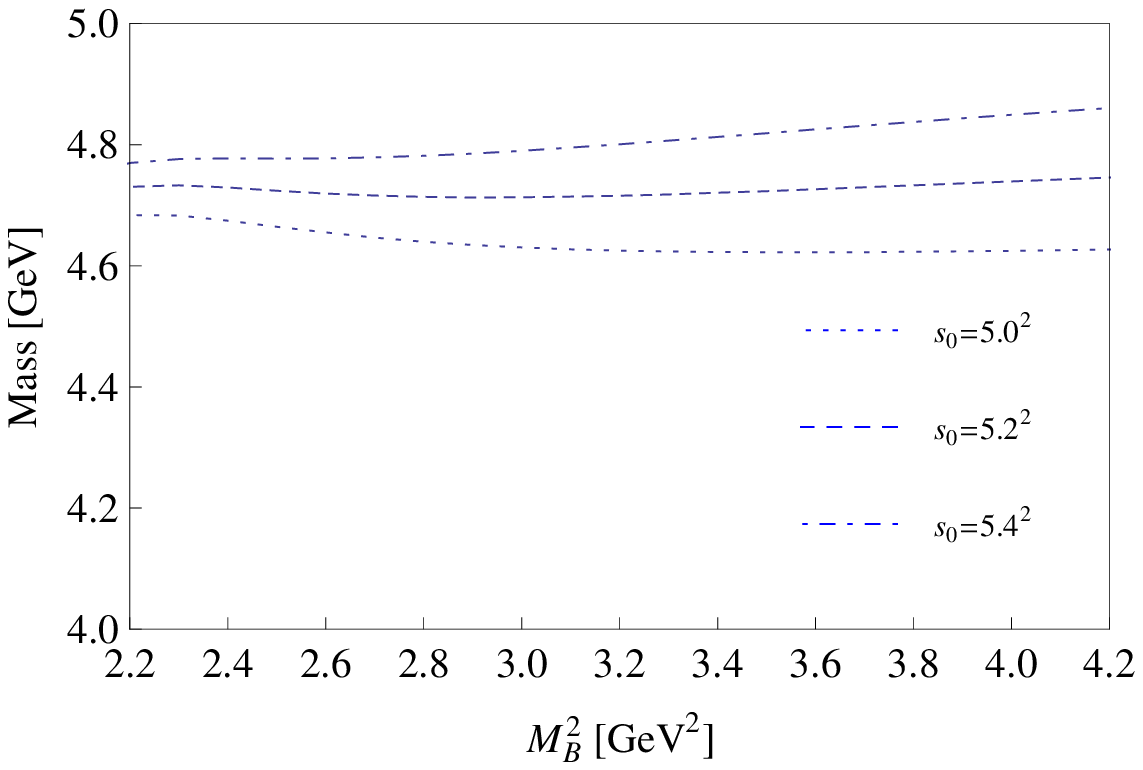}}&
\scalebox{0.66}{\includegraphics{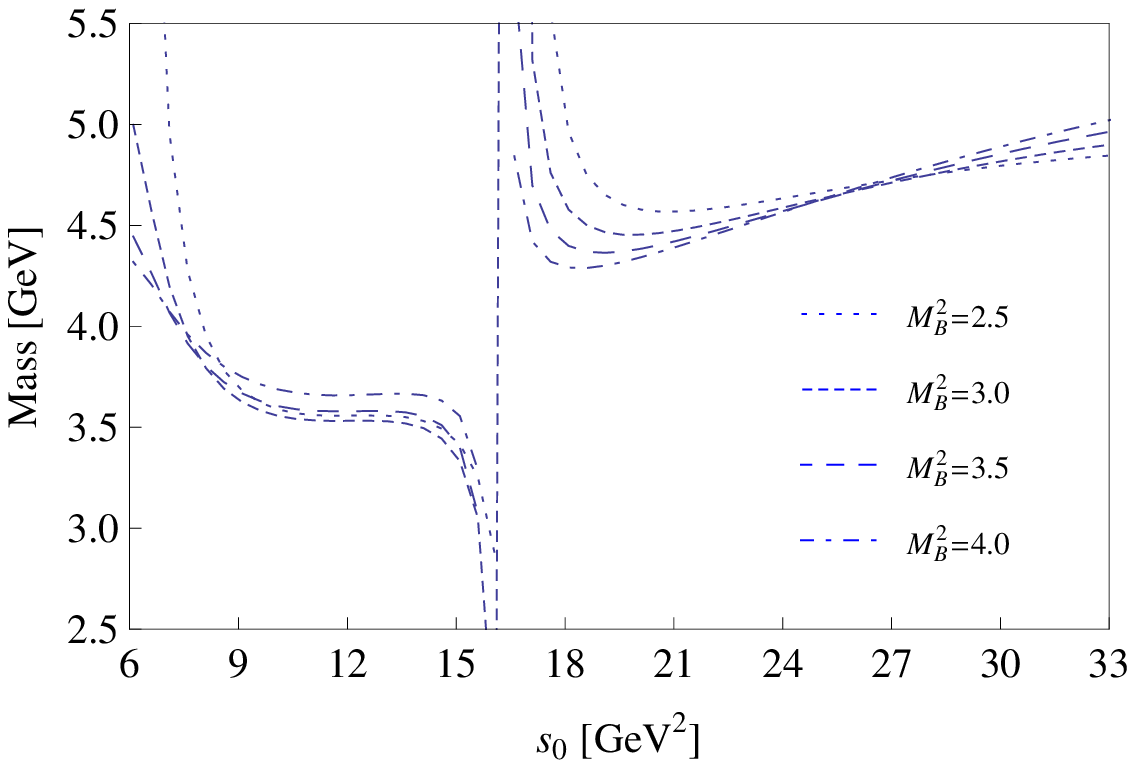}}
\end{tabular}
\figcaption{The variation of $M_X$ with $M^2_B$ (left) and $s_0$
(right) for the current $\eta_{6}$.} \label{mass6}
\end{center}

Replacing $m_c$ with $m_b$ and repeating the same analysis
procedures done above, we collect the results of the
bottomonium-like systems in Table \ref{table2}.
%
\begin{center}
\begin{tabular}{c|c|c|c|c|c}
\hline \hline
        & Currents & $s_0(\mbox{GeV}^2)$ & $[M^2_{\mbox{min}}$, $M^2_{\mbox{max}}](\mbox{GeV}^2)$& $m_{X_b}$\mbox{(GeV)}& PC(\%)\\
\hline           & $\eta_{1b}$      &  $11.2^2$            & $6.4\sim9.4$                 & $-$              & - \\
                 & $\eta_{2b}$      &  $11.2^2$            & $6.4\sim9.5$                 & $10.64\pm0.12$    & 45.2  \\
$J^{PC}=0^{--}$  & $\eta_{3b}$      &  $11.2^2$            & $6.4\sim9.5$                 & $-$              & - \\
                 & $\eta_{4b}$      &  $11.2^2$            & $6.4\sim9.5$                 & $10.64\pm0.12$    & 45.1 \\
\hline
                 & $\eta_{5b}$      &  $11.2^2$            & $6.4\sim9.4$                 & $-$              & - \\
                 & $\eta_{6b}$      &  $11.2^2$            & $6.4\sim9.4$                 & $10.67\pm0.11$    & 44.2 \\
                 & $\eta_{7b}$      &  $11.2^2$            & $6.4\sim9.7$                 & $-$              & - \\
$J^{PC}=0^{-+}$  & $\eta_{8b}$      &  $11.2^2$            & $6.4\sim9.6$                 & $-$              & - \\
                 & $\eta_{9b}$      &  $11.2^2$            & $6.4\sim9.5$                 & $10.64\pm0.12$    & 45.1 \\
                 & $\eta_{10b}$     &  $11.2^2$            & $6.4\sim9.5$                 & $10.64\pm0.11$    & 45.6 \\
\hline \hline
\end{tabular}
\tabcaption{The threshold value, Borel window, mass and pole
contribution for $\eta_{1b}\sim\eta_{10b}$. The mass and pole
contribution are calculated at $M_B^2=9.0$GeV$^2$. \label{table2}}
\end{center}

\section{Summary}\label{sec:summary}

We have constructed the charmonium-like tetraquark interpolating
currents with $J^{PC}=0^{--}, 0^{-+}$ using the
diquark-antidiquark fields. Then we calculated the correlation
functions and the spectral densities of these currents. Both the
dimension 3 quark condensate and dimension 5 quark gluon mixed
condensate vanish if we take $m_{u,d}=0$. The special Lorenz
structures of the currents prohibit their appearance in the OPE.
The four quark condensate $\langle\bar qq\rangle^2$ becomes the
most important power correction numerically. It is much bigger
than the gluon condensates $\langle g^2GG\rangle$, $\langle
g^3fGGG\rangle$ and the dimension 8 condensate $\langle\bar
qg\sigma\cdot Gq\rangle\langle\bar qq\rangle$.

In the working region of the Borel parameter, the variation of the
extracted mass with $s_0$ and $M_B$ is stable for the currents
$\eta_2, \eta_4, \eta_6, \eta_9, \eta_{10}$. For the $0^{--}$
charmonium-like states, its mass is around $4.5$ GeV. For the
$0^{-+}$ charmonium-like state, the mass is around $4.6$ GeV. For
the $0^{--}$ and $0^{-+}$ bottomonium-like states, their masses are
around $10.6$ GeV. It's interesting to note that the extracted mass
value $\sim 4.5$ GeV of the $0^{--}$ charmonium-like state is quite
close to the mass value $4.1 \sim 4.4$ GeV in Ref.
\cite{1980-Chao-p281-281}.

The possible decay modes of the $0^{--}$ charmonium-like state are
straightforward after making Fierz transformation to the diquark
type interpolating currents in Eq. \ref{currents1}. They can be
expressed in terms of the linear combination of the meson-meson type
of operators such as:
$(\bar{q}_{a}\gamma_{\mu}q_{a})(\bar{c}_{b}\gamma^{\mu}\gamma_5
c_{b}),
(\bar{q}_{a}\gamma_{\mu}\gamma_{5}q_{a})(\bar{c}_{b}\gamma^{\mu}c_{b}),
(\bar{q}_{a}\gamma_{\mu}c_{a})(\bar{c}_{b}\gamma^{\mu}\gamma_{5}q_{b})
 +(\bar{q}_{a}\gamma_{\mu}\gamma_{5}c_{a})(\bar{c}_{b}\gamma^{\mu}q_{b}),
(\bar{q}_{a}c_{a})(\bar{c}_{b}\gamma_5q_{b})-(\bar{q}_{a}\gamma_5c_{a})(\bar{c}_{b}q_{b})
$.

There are two types of the $0^{--}$ charmonium-like state with
different isospin and G-parity: $I^G=1^+$ and $I^G=0^-$. Considering
the conservation of the isospin, G-parity and C parity, we collect
the S-wave and P-wave decay modes of the possible $0^{--}$
charmonium-like state in Table \ref{table3}.

Clearly the S-wave decay modes are dominant. The S-wave decay
products always contain a P-wave and S-wave meson pair. Such a decay
pattern is also characteristic of the hybrid meson. Although the
exotic $0^{--}$ state can not be composed of a $c\bar c$ pair, it
can be a $cG\bar c$ hybrid state. In fact, the $0^{--}$
charmonium-like tetraquark operator and the $0^{--}$ $cG\bar c$
hybrid operator probably couple to the same $0^{--}$ physical state.

This interesting $0^{--}$ state may be searched for experimentally
at facilities such as Super-B factories, PANDE, LHC and RHIC in the
future. Especially at RHIC, plenty of charm, anti-charm and light
quarks are produced simultaneously which may be helpful to the
formation of the $0^{--}$ charmonium-like state.

\begin{center}
\begin{tabular}{c|c|c}
\hline \hline
        $I^G$  &                     $S$-wave                       &                           $P$-wave               \\
\hline         &    $D^{\ast}(2007)^0\bar{D}_1(2420)^0+c.c.$,       &  $D^0(1865)\bar{D}^{\ast}(2007)^0+c.c.$,\\
        $0^-$  &    $D_0^{\ast}(2400)^0\bar{D}^0(1865)+c.c.$,       &  $D^{\ast}(2007)^0\bar{D}^{\ast}(2007)^0$,  \\
               &    $\omega(782)\chi_{c1}(1P)$, $J/\psi f_1(1285)$  &  $J/\psi\eta$, $J/\psi\eta^{\prime}$, $\psi(2S)\eta$, $\eta_c(1S)\omega$,\\
               &                                                    &  $\eta_c(2S)\omega$, $h_c(1P)\sigma$, $h_c(1P)f_0(980)$\\
\hline         &    $D^{\ast}(2007)^0\bar{D}_1(2420)^0+c.c.$,       &  $D^0(1865)\bar{D}^{\ast}(2007)^0+c.c.$,\\
        $1^+$  &    $D_0^{\ast}(2400)^0\bar{D}^0(1865)+c.c.$,       &  $D^{\ast}(2007)^0\bar{D}^{\ast}(2007)^0$,  \\
               &    $\rho(770)\chi_{c1}(1P)$, $J/\psi a_1(1260)$    &  $\eta_c(1S)\rho$, $\eta_c(2S)\rho$, $h_c(1P)a_0(980)$ \\
               &                                                    &  $J/\psi\pi$, $J/\psi\pi_1(1400)$, $\psi(2S)\pi$\\
\hline \hline
\end{tabular}
\tabcaption{The possible decay modes of the $0^{--}$
charmonium-like state. \label{table3}}
\end{center}

\section*{Acknowledgments}
The authors thank Professor W. Z. Deng for useful discussions. W.
Chen is grateful for J. R. Zhang, M.Q. Huang and M. E. Bracco for
their helpful email discussions. This project was supported by the
National Natural Science Foundation of China under Grants 10625521,
10721063 and Ministry of Science and Technology of China
(2009CB825200).

%



\appendix
\section{THE SPECTRAL DENSITIES}
In this appendix we show the spectral densities of the tetraquark
interpolating currents defined in
Eqs.~(\ref{currents1})-(\ref{currents2}). The same subscripts are
used to denote the results for the currents $\eta_1, \eta_2, \eta_3,
\eta_4, \eta_5, \eta_6, \eta_7, \eta_8, \eta_9, \eta_{10}$:
\begin{eqnarray}
\rho^{OPE}=\rho^{pert}(s)+\rho^{\langle\bar qq\rangle}(s)+\rho^{\langle G^2\rangle}(s)+\rho^{mix}(s)
+\rho^{\langle\bar qq\rangle^2}(s) \label{rhopart}
\end{eqnarray}
For the expressions below, the integration limits are:
\begin{eqnarray}
\nonumber \alpha_{max}&=&\frac{1+\sqrt{1-4m_c^2/s}}{2},\hspace{1cm} \alpha_{min}=\frac{1-\sqrt{1-4m_c^2/s}}{2}\\
\nonumber \beta_{max}&=&1-\alpha,\hspace{2.8cm} \beta_{min}=\frac{\alpha m_c^2}{\alpha s-m_c^2}.
\end{eqnarray}


\begin{enumerate}
\item The spectral densities of the interpolating currents with the quantum numbers $J^{PC}=0^{--}:$   \label{rho0--}
\\

For $\eta_1$:
\begin{eqnarray}
\nonumber
\rho^{pert}_1(s)&=&\frac{1}{2^7\pi^6}\int^{\alpha_{max}}_{\alpha_{min}}\frac{d\alpha}{\alpha^3}\int^{\beta_{max}}_{\beta_{min}}
\frac{d\beta}{\beta^3}(1-\alpha-\beta)^2[(\alpha+\beta)m_c^2-3\alpha\beta
s][(\alpha+\beta)m_c^2-\alpha\beta s]^3 \, ,
\\ \nonumber
\rho^{\langle\bar qq\rangle}_1(s)&=&0 \, ,
\\ \nonumber
\rho^{\langle GG\rangle}_1(s)&=&\frac{\langle
g_s^2GG\rangle}{2^7\pi^6}\int^{\alpha_{max}}_{\alpha_{min}}\frac{d\alpha}{\alpha^2}\int^{\beta_{max}}_{\beta_{min}}
d\beta\{
\frac{(1-\alpha-\beta)^2m_c^2}{3\alpha}[2(\alpha+\beta)m_c^2-3\alpha\beta
s]
\\ \nonumber
&&\hspace{4cm}-\frac{1-\alpha-\beta}{2\beta}[(\alpha+\beta)m_c^2-2\alpha\beta s][(\alpha+\beta)m_c^2-\alpha\beta s]\} \, ,
\\ \nonumber
\rho^{mix}_1(s)&=&0 \, ,
\\
\rho^{\langle\bar qq\rangle^2}_1(s)&=&-\frac{m_c^2\langle\bar qq\rangle^2}{3\pi^2}\sqrt{1-4m_c^2/s} \, ,
\end{eqnarray}
\begin{eqnarray}
\nonumber
\Pi^{mix\langle\bar qq\rangle}_1(M_B^2)&=&-\frac{m_c^2\langle\bar qg_s\sigma\cdot Gq\rangle\langle\bar qq\rangle}{12\pi^2}\int_0^1
\frac{d\alpha}{\alpha}(\frac{2m_c^2}{\alpha M_B^2}+1)e^{-\frac{m_c^2}{\alpha(1-\alpha)M_B^2}} \, ,
\\ \nonumber
\Pi^{\langle GGG\rangle}_1(M_B^2)&=&-\frac{\langle g_s^3fGGG\rangle}{3\times2^8\pi^6}\int_0^1d\alpha\int_0^{\beta_{max}}d\beta
\\ \nonumber
&& \{\frac{\ln(\alpha\beta(1-\alpha-\beta)M_B^4)-2\ln2-\gamma_E}{\alpha\beta}[12(\alpha\beta M_B^2)^2
+6\alpha\beta M_B^2(\alpha+\beta)m_c^2
+(\alpha+\beta)^2m_c^4]
\\ \nonumber
&&+\frac{33(\alpha\beta M_B^2)^2+12\alpha\beta M_B^2(\alpha+\beta)m_c^2+(\alpha+\beta)^2m_c^4}{\alpha\beta}
+\frac{(1-\alpha-\beta)^2m_c^2}{\alpha^4}[2\alpha\beta M_B^2+(\alpha+\beta)m_c^2]
\\ \nonumber
&&-\frac{(1-\alpha-\beta)^2}{2\alpha^3}[3\alpha\beta M_B^4+M_B^2(\alpha+\beta)m_c^2]\}e^{-\frac{(\alpha+\beta)m_c^2}{\alpha\beta M_B^2}} \, .
\end{eqnarray}
For $\eta_2$:
\begin{eqnarray}
\nonumber
\rho^{pert}_2(s)&=&\frac{1}{2^5\pi^6}\int^{\alpha_{max}}_{\alpha_{min}}\frac{d\alpha}{\alpha^3}\int^{\beta_{max}}_{\beta_{min}}
\frac{d\beta}{\beta^3}(1-\alpha-\beta)^2[(\alpha+\beta)m_c^2-3\alpha\beta
s][(\alpha+\beta)m_c^2-\alpha\beta s]^3 \, ,
\\ \nonumber
\rho^{\langle\bar qq\rangle}_2(s)&=&0 \, ,
\\ \nonumber
\rho^{\langle GG\rangle}_2(s)&=&\frac{\langle
g_s^2GG\rangle}{2^5\pi^6}\int^{\alpha_{max}}_{\alpha_{min}}\frac{d\alpha}{\alpha^2}\int^{\beta_{max}}_{\beta_{min}}
d\beta\{
\frac{(1-\alpha-\beta)^2m_c^2}{3\alpha}[2(\alpha+\beta)m_c^2-3\alpha\beta
s]
\\ \nonumber
&&\hspace{4cm}+\frac{5(1-\alpha-\beta)}{4\beta}[(\alpha+\beta)m_c^2-2\alpha\beta s][(\alpha+\beta)m_c^2-\alpha\beta s]\} \, ,
\\ \nonumber
\rho^{mix}_2(s)&=&0 \, ,
\\
\rho^{\langle\bar qq\rangle^2}_2(s)&=&-\frac{4m_c^2\langle\bar qq\rangle^2}{3\pi^2}\sqrt{1-4m_c^2/s} \, ,
\end{eqnarray}
\begin{eqnarray}
\nonumber
\Pi^{mix\langle\bar qq\rangle}_2(M_B^2)&=&-\frac{m_c^2\langle\bar qg_s\sigma\cdot Gq\rangle\langle\bar qq\rangle}{6\pi^2}\int_0^1
\frac{d\alpha}{\alpha}(\frac{4m_c^2}{\alpha M_B^2}-5)e^{-\frac{m_c^2}{\alpha(1-\alpha)M_B^2}} \, ,
\\ \nonumber
\Pi^{\langle GGG\rangle}_2(M_B^2)&=&-\frac{\langle g_s^3fGGG\rangle}{3\times2^6\pi^6}\int_0^1d\alpha\int_0^{\beta_{max}}d\beta
\\ \nonumber
&& \{\frac{\ln(\alpha\beta(1-\alpha-\beta)M_B^4)-2\ln2-\gamma_E}{\alpha\beta}[12(\alpha\beta M_B^2)^2
+6\alpha\beta M_B^2(\alpha+\beta)m_c^2
+(\alpha+\beta)^2m_c^4]
\\ \nonumber
&&+\frac{33(\alpha\beta M_B^2)^2+12\alpha\beta M_B^2(\alpha+\beta)m_c^2+(\alpha+\beta)^2m_c^4}{\alpha\beta}
+\frac{(1-\alpha-\beta)^2m_c^2}{\alpha^4}[2\alpha\beta M_B^2+(\alpha+\beta)m_c^2]
\\ \nonumber
&&-\frac{(1-\alpha-\beta)^2}{2\alpha^3}[3\alpha\beta M_B^4+M_B^2(\alpha+\beta)m_c^2]\}e^{-\frac{(\alpha+\beta)m_c^2}{\alpha\beta M_B^2}} \, .
\end{eqnarray}
For $\eta_3$:
\begin{eqnarray}
\nonumber
\rho^{pert}_3(s)&=&\frac{1}{2^6\pi^6}\int^{\alpha_{max}}_{\alpha_{min}}\frac{d\alpha}{\alpha^3}\int^{\beta_{max}}_{\beta_{min}}
\frac{d\beta}{\beta^3}(1-\alpha-\beta)^2[(\alpha+\beta)m_c^2-3\alpha\beta
s][(\alpha+\beta)m_c^2-\alpha\beta s]^3 \, ,
\\ \nonumber
\rho^{\langle\bar qq\rangle}_3(s)&=&0 \, ,
\\ \nonumber
\rho^{\langle GG\rangle}_3(s)&=&\frac{\langle
g_s^2GG\rangle}{2^6\pi^6}\int^{\alpha_{max}}_{\alpha_{min}}\frac{d\alpha}{\alpha^2}\int^{\beta_{max}}_{\beta_{min}}
d\beta\{
\frac{(1-\alpha-\beta)^2m_c^2}{3\alpha}[2(\alpha+\beta)m_c^2-3\alpha\beta
s]
\\ \nonumber
&&\hspace{4cm}+\frac{1-\alpha-\beta}{2\beta}[(\alpha+\beta)m_c^2-2\alpha\beta s][(\alpha+\beta)m_c^2-\alpha\beta s]\} \, ,
\\ \nonumber
\rho^{mix}_3(s)&=&0 \, ,
\\
\rho^{\langle\bar qq\rangle^2}_3(s)&=&-\frac{2m_c^2\langle\bar qq\rangle^2}{3\pi^2}\sqrt{1-4m_c^2/s} \, ,
\end{eqnarray}
\begin{eqnarray}
\nonumber
\Pi^{mix\langle\bar qq\rangle}_3(M_B^2)&=&-\frac{m_c^2\langle\bar qg_s\sigma\cdot Gq\rangle\langle\bar qq\rangle}{6\pi^2}\int_0^1
\frac{d\alpha}{\alpha}(\frac{2m_c^2}{\alpha M_B^2}-1)e^{-\frac{m_c^2}{\alpha(1-\alpha)M_B^2}} \, ,
\\ \nonumber
\Pi^{\langle GGG\rangle}_3(M_B^2)&=&-\frac{\langle g_s^3fGGG\rangle}{3\times2^6\pi^6}\int_0^1d\alpha\int_0^{\beta_{max}}d\beta
\\ \nonumber
&& \{\frac{\ln(\alpha\beta(1-\alpha-\beta)M_B^4)-2\ln2-\gamma_E}{\alpha\beta}[12(\alpha\beta M_B^2)^2
+6\alpha\beta M_B^2(\alpha+\beta)m_c^2
+(\alpha+\beta)^2m_c^4]
\\ \nonumber
&&+\frac{33(\alpha\beta M_B^2)^2+12\alpha\beta M_B^2(\alpha+\beta)m_c^2+(\alpha+\beta)^2m_c^4}{\alpha\beta}
+\frac{(1-\alpha-\beta)^2m_c^2}{\alpha^4}[2\alpha\beta M_B^2+(\alpha+\beta)m_c^2]
\\ \nonumber
&&-\frac{(1-\alpha-\beta)^2}{2\alpha^3}[3\alpha\beta M_B^4+M_B^2(\alpha+\beta)m_c^2]\}e^{-\frac{(\alpha+\beta)m_c^2}{\alpha\beta M_B^2}} \, .
\end{eqnarray}
For $\eta_4$:
\begin{eqnarray}
\nonumber
\rho^{pert}_4(s)&=&\frac{1}{2^8\pi^6}\int^{\alpha_{max}}_{\alpha_{min}}\frac{d\alpha}{\alpha^3}\int^{\beta_{max}}_{\beta_{min}}
\frac{d\beta}{\beta^3}(1-\alpha-\beta)^2[(\alpha+\beta)m_c^2-3\alpha\beta
s][(\alpha+\beta)m_c^2-\alpha\beta s]^3 \, ,
\\ \nonumber
\rho^{\langle\bar qq\rangle}_4(s)&=&0 \, ,
\\ \nonumber
\rho^{\langle GG\rangle}_4(s)&=&\frac{\langle
g_s^2GG\rangle}{2^8\pi^6}\int^{\alpha_{max}}_{\alpha_{min}}\frac{d\alpha}{\alpha^2}\int^{\beta_{max}}_{\beta_{min}}
d\beta\{
\frac{(1-\alpha-\beta)^2m_c^2}{3\alpha}[2(\alpha+\beta)m_c^2-3\alpha\beta
s]
\\ \nonumber
&&\hspace{4cm}+\frac{1-\alpha-\beta}{\beta}[(\alpha+\beta)m_c^2-2\alpha\beta s][(\alpha+\beta)m_c^2-\alpha\beta s]\} \, ,
\\ \nonumber
\rho^{mix}_4(s)&=&0 \, ,
\\
\rho^{\langle\bar qq\rangle^2}_4(s)&=&-\frac{m_c^2\langle\bar qq\rangle^2}{6\pi^2}\sqrt{1-4m_c^2/s} \, ,
\end{eqnarray}
\begin{eqnarray}
\nonumber
\Pi^{mix\langle\bar qq\rangle}_4(M_B^2)&=&-\frac{m_c^2\langle\bar qg_s\sigma\cdot Gq\rangle\langle\bar qq\rangle}{12\pi^2}\int_0^1
\frac{d\alpha}{\alpha}(\frac{m_c^2}{\alpha M_B^2}-1)e^{-\frac{m_c^2}{\alpha(1-\alpha)M_B^2}} \, ,
\\ \nonumber
\Pi^{\langle GGG\rangle}_4(M_B^2)&=&-\frac{\langle g_s^3fGGG\rangle}{3\times2^9\pi^6}\int_0^1d\alpha\int_0^{\beta_{max}}d\beta
\\ \nonumber
&& \{\frac{\ln(\alpha\beta(1-\alpha-\beta)M_B^4)-2\ln2-\gamma_E}{\alpha\beta}[12(\alpha\beta M_B^2)^2
+6\alpha\beta M_B^2(\alpha+\beta)m_c^2
+(\alpha+\beta)^2m_c^4]
\\ \nonumber
&&+\frac{33(\alpha\beta M_B^2)^2+12\alpha\beta M_B^2(\alpha+\beta)m_c^2+(\alpha+\beta)^2m_c^4}{\alpha\beta}
+\frac{(1-\alpha-\beta)^2m_c^2}{\alpha^4}[2\alpha\beta M_B^2+(\alpha+\beta)m_c^2]
\\ \nonumber
&&-\frac{(1-\alpha-\beta)^2}{2\alpha^3}[3\alpha\beta M_B^4+M_B^2(\alpha+\beta)m_c^2]\}e^{-\frac{(\alpha+\beta)m_c^2}{\alpha\beta M_B^2}} \, .
\end{eqnarray}


\item The spectral densities of the interpolating currents with the quantum numbers $J^{PC}=0^{-+}:$   \label{rho0-+}
\\

For $\eta_5$:
\begin{eqnarray}
\nonumber
\rho^{pert}_5(s)&=&\frac{1}{2^7\pi^6}\int^{\alpha_{max}}_{\alpha_{min}}\frac{d\alpha}{\alpha^3}\int^{\beta_{max}}_{\beta_{min}}
\frac{d\beta}{\beta^3}(1-\alpha-\beta)^2[(\alpha+\beta)m_c^2-3\alpha\beta
s][(\alpha+\beta)m_c^2-\alpha\beta s]^3 \, ,
\\ \nonumber
\rho^{\langle\bar qq\rangle}_5(s)&=&0 \, ,
\\ \nonumber
\rho^{\langle GG\rangle}_5(s)&=&\frac{\langle
g_s^2GG\rangle}{2^7\pi^6}\int^{\alpha_{max}}_{\alpha_{min}}\frac{d\alpha}{\alpha^2}\int^{\beta_{max}}_{\beta_{min}}
d\beta\{
\frac{(1-\alpha-\beta)^2m_c^2}{3\alpha}[2(\alpha+\beta)m_c^2-3\alpha\beta
s]
\\ \nonumber
&&\hspace{4cm}-\frac{1-\alpha-\beta}{2\beta}[(\alpha+\beta)m_c^2-2\alpha\beta s][(\alpha+\beta)m_c^2-\alpha\beta s]\} \, ,
\\ \nonumber
\rho^{mix}_5(s)&=&0 \, ,
\\
\rho^{\langle\bar qq\rangle^2}_5(s)&=&-\frac{m_c^2\langle\bar qq\rangle^2}{3\pi^2}\sqrt{1-4m_c^2/s} \, ,
\end{eqnarray}
\begin{eqnarray}
\nonumber
\Pi^{mix\langle\bar qq\rangle}_5(M_B^2)&=&-\frac{m_c^2\langle\bar qg_s\sigma\cdot Gq\rangle\langle\bar qq\rangle}{12\pi^2}\int_0^1
\frac{d\alpha}{\alpha}(\frac{2m_c^2}{\alpha M_B^2}+1)e^{-\frac{m_c^2}{\alpha(1-\alpha)M_B^2}} \, ,
\\ \nonumber
\Pi^{\langle GGG\rangle}_5(M_B^2)&=&-\frac{\langle g_s^3fGGG\rangle}{3\times2^8\pi^6}\int_0^1d\alpha\int_0^{\beta_{max}}d\beta
\\ \nonumber
&& \{\frac{\ln(\alpha\beta(1-\alpha-\beta)M_B^4)-2\ln2-\gamma_E}{\alpha\beta}[12(\alpha\beta M_B^2)^2
+6\alpha\beta M_B^2(\alpha+\beta)m_c^2
+(\alpha+\beta)^2m_c^4]
\\ \nonumber
&&+\frac{33(\alpha\beta M_B^2)^2+12\alpha\beta M_B^2(\alpha+\beta)m_c^2+(\alpha+\beta)^2m_c^4}{\alpha\beta}
+\frac{(1-\alpha-\beta)^2m_c^2}{\alpha^4}[2\alpha\beta M_B^2+(\alpha+\beta)m_c^2]
\\ \nonumber
&&-\frac{(1-\alpha-\beta)^2}{2\alpha^3}[3\alpha\beta M_B^4+M_B^2(\alpha+\beta)m_c^2]\}e^{-\frac{(\alpha+\beta)m_c^2}{\alpha\beta M_B^2}} \, .
\end{eqnarray}
For $\eta_6$:
\begin{eqnarray}
\nonumber
\rho^{pert}_6(s)&=&\frac{3}{2^6\pi^6}\int^{\alpha_{max}}_{\alpha_{min}}\frac{d\alpha}{\alpha^3}\int^{\beta_{max}}_{\beta_{min}}
\frac{d\beta}{\beta^3}(1-\alpha-\beta)^2[(\alpha+\beta)m_c^2-3\alpha\beta
s][(\alpha+\beta)m_c^2-\alpha\beta s]^3 \, ,
\\ \nonumber
\rho^{\langle\bar qq\rangle}_6(s)&=&0 \, ,
\\ \nonumber
\rho^{\langle GG\rangle}_6(s)&=&\frac{\langle
g_s^2GG\rangle}{2^6\pi^6}\int^{\alpha_{max}}_{\alpha_{min}}\frac{d\alpha}{\alpha^2}\int^{\beta_{max}}_{\beta_{min}}
d\beta\{
\frac{(1-\alpha-\beta)^2m_c^2}{\alpha}[2(\alpha+\beta)m_c^2-3\alpha\beta
s]
\\ \nonumber
&&\hspace{4cm}+\frac{(1-\alpha-\beta)^2+2\alpha\beta}{4\beta^2}[(\alpha+\beta)m_c^2-2\alpha\beta s][(\alpha+\beta)m_c^2-\alpha\beta s] \, ,
\\ \nonumber
\rho^{mix}_6(s)&=&0 \, ,
\\
\rho^{\langle\bar qq\rangle^2}_6(s)&=&-\frac{2m_c^2\langle\bar qq\rangle^2}{\pi^2}\sqrt{1-4m_c^2/s} \, ,
\end{eqnarray}
\begin{eqnarray}
\nonumber
\Pi^{mix\langle\bar qq\rangle}_6(M_B^2)&=&-\frac{m_c^2\langle\bar qg_s\sigma\cdot Gq\rangle\langle\bar qq\rangle}{\pi^2}\int_0^1
\frac{d\alpha}{\alpha^2}\frac{m_c^2}{M_B^2}e^{-\frac{m_c^2}{\alpha(1-\alpha)M_B^2}} \, ,
\\ \nonumber
\Pi^{\langle GGG\rangle}_6(M_B^2)&=&-\frac{\langle g_s^3fGGG\rangle}{3\times2^7\pi^6}\int_0^1d\alpha\int_0^{\beta_{max}}d\beta
\\ \nonumber
&&\{\frac{\ln(\alpha\beta(1-\alpha-\beta)M_B^4)-2\ln2-\gamma_E}{\alpha\beta}[12(\alpha\beta M_B^2)^2
+6\alpha\beta M_B^2(\alpha+\beta)m_c^2+(\alpha+\beta)^2m_c^4]
\\ \nonumber
&&+\frac{33(\alpha\beta M_B^2)^2+12\alpha\beta M_B^2(\alpha+\beta)m_c^2+(\alpha+\beta)^2m_c^4}{\alpha\beta}
+\frac{3(1-\alpha-\beta)^2m_c^2}{\alpha^4}[2\alpha\beta M_B^2+(\alpha+\beta)m_c^2]
\\ \nonumber
&&-\frac{3(1-\alpha-\beta)^2}{2\alpha^3}[3\alpha\beta M_B^4+M_B^2(\alpha+\beta)m_c^2]
+\frac{2}{1-\alpha-\beta}[3\alpha\beta M_B^4+M_B^2(\alpha+\beta)m_c^2]\}e^{-\frac{(\alpha+\beta)m_c^2}{\alpha\beta M_B^2}} \, .
\end{eqnarray}
For $\eta_7$:
\begin{eqnarray}
\nonumber
\rho^{pert}_7(s)&=&\frac{1}{2^5\pi^6}\int^{\alpha_{max}}_{\alpha_{min}}\frac{d\alpha}{\alpha^3}\int^{\beta_{max}}_{\beta_{min}}
\frac{d\beta}{\beta^3}(1-\alpha-\beta)^2[(\alpha+\beta)m_c^2-3\alpha\beta
s][(\alpha+\beta)m_c^2-\alpha\beta s]^3 \, ,
\\ \nonumber
\rho^{\langle\bar qq\rangle}_7(s)&=&0 \, ,
\\ \nonumber
\rho^{\langle GG\rangle}_7(s)&=&\frac{\langle
g_s^2GG\rangle}{3\times2^5\pi^6}\int^{\alpha_{max}}_{\alpha_{min}}\frac{d\alpha}{\alpha^2}\int^{\beta_{max}}_{\beta_{min}}
d\beta\{
\frac{(1-\alpha-\beta)^2m_c^2}{\alpha}[2(\alpha+\beta)m_c^2-3\alpha\beta
s]
\\ \nonumber
&&\hspace{4cm}+\frac{15(1-\alpha-\beta)^2+30\alpha\beta}{16\beta^2}[(\alpha+\beta)m_c^2-2\alpha\beta s][(\alpha+\beta)m_c^2-\alpha\beta s]\} \, ,
\\ \nonumber
\rho^{mix}_7(s)&=&0 \, ,
\\
\rho^{\langle\bar qq\rangle^2}_7(s)&=&-\frac{4m_c^2\langle\bar qq\rangle^2}{3\pi^2}\sqrt{1-4m_c^2/s} \, ,
\end{eqnarray}
\begin{eqnarray}
\nonumber
\Pi^{mix\langle\bar qq\rangle}_7(M_B^2)&=&-\frac{2m_c^2\langle\bar qg_s\sigma\cdot Gq\rangle\langle\bar qq\rangle}{3\pi^2}\int_0^1
\frac{d\alpha}{\alpha^2}\frac{m_c^2}{M_B^2}e^{-\frac{m_c^2}{\alpha(1-\alpha)M_B^2}} \, ,
\\ \nonumber
\Pi^{\langle GGG\rangle}_7(M_B^2)&=&-\frac{\langle g_s^3fGGG\rangle}{3\times2^6\pi^6}\int_0^1d\alpha\int_0^{\beta_{max}}d\beta
\\ \nonumber
&& \{\frac{\ln(\alpha\beta(1-\alpha-\beta)M_B^4)-2\ln2-\gamma_E}{\alpha\beta}[12(\alpha\beta M_B^2)^2
+6\alpha\beta M_B^2(\alpha+\beta)m_c^2
+(\alpha+\beta)^2m_c^4]
\\ \nonumber
&&+\frac{33(\alpha\beta M_B^2)^2+12\alpha\beta M_B^2(\alpha+\beta)m_c^2+(\alpha+\beta)^2m_c^4}{\alpha\beta}
+\frac{(1-\alpha-\beta)^2m_c^2}{\alpha^4}[2\alpha\beta M_B^2+(\alpha+\beta)m_c^2]
\\ \nonumber
&&-\frac{(1-\alpha-\beta)^2}{2\alpha^3}[3\alpha\beta M_B^4+M_B^2(\alpha+\beta)m_c^2]\}e^{-\frac{(\alpha+\beta)m_c^2}{\alpha\beta M_B^2}} \, .
\end{eqnarray}
For $\eta_8$:
\begin{eqnarray}
\nonumber
\rho^{pert}_8(s)&=&\frac{1}{2^6\pi^6}\int^{\alpha_{max}}_{\alpha_{min}}\frac{d\alpha}{\alpha^3}\int^{\beta_{max}}_{\beta_{min}}
\frac{d\beta}{\beta^3}(1-\alpha-\beta)^2[(\alpha+\beta)m_c^2-3\alpha\beta
s][(\alpha+\beta)m_c^2-\alpha\beta s]^3 \, ,
\\ \nonumber
\rho^{\langle\bar qq\rangle}_8(s)&=&0 \, ,
\\ \nonumber
\rho^{\langle GG\rangle}_8(s)&=&\frac{\langle
g_s^2GG\rangle}{3\times2^6\pi^6}\int^{\alpha_{max}}_{\alpha_{min}}\frac{d\alpha}{\alpha^2}\int^{\beta_{max}}_{\beta_{min}}
d\beta\{
\frac{(1-\alpha-\beta)^2m_c^2}{\alpha}[2(\alpha+\beta)m_c^2-3\alpha\beta
s]
\\ \nonumber
&&\hspace{4cm}+\frac{3(1-\alpha-\beta)^2+6\alpha\beta}{8\beta^2}[(\alpha+\beta)m_c^2-2\alpha\beta s][(\alpha+\beta)m_c^2-\alpha\beta s]\} \, ,
\\ \nonumber
\rho^{mix}_8(s)&=&0 \, ,
\\
\rho^{\langle\bar qq\rangle^2}_8(s)&=&-\frac{2m_c^2\langle\bar qq\rangle^2}{3\pi^2}\sqrt{1-4m_c^2/s} \, ,
\end{eqnarray}
\begin{eqnarray}
\nonumber
\Pi^{mix\langle\bar qq\rangle}_8(M_B^2)&=&-\frac{m_c^2\langle\bar qg_s\sigma\cdot Gq\rangle\langle\bar qq\rangle}{3\pi^2}\int_0^1
\frac{d\alpha}{\alpha^2}\frac{m_c^2}{M_B^2}e^{-\frac{m_c^2}{\alpha(1-\alpha)M_B^2}} \, ,
\\ \nonumber
\Pi^{\langle GGG\rangle}_8(M_B^2)&=&-\frac{\langle g_s^3fGGG\rangle}{3\times2^7\pi^6}\int_0^1d\alpha\int_0^{\beta_{max}}d\beta
\\ \nonumber
&& \{\frac{\ln(\alpha\beta(1-\alpha-\beta)M_B^4)-2\ln2-\gamma_E}{\alpha\beta}[12(\alpha\beta M_B^2)^2
+6\alpha\beta M_B^2(\alpha+\beta)m_c^2
+(\alpha+\beta)^2m_c^4]
\\ \nonumber
&&+\frac{33(\alpha\beta M_B^2)^2+12\alpha\beta M_B^2(\alpha+\beta)m_c^2+(\alpha+\beta)^2m_c^4}{\alpha\beta}
+\frac{(1-\alpha-\beta)^2m_c^2}{\alpha^4}[2\alpha\beta M_B^2+(\alpha+\beta)m_c^2]
\\ \nonumber
&&-\frac{(1-\alpha-\beta)^2}{2\alpha^3}[3\alpha\beta M_B^4+M_B^2(\alpha+\beta)m_c^2]\}e^{-\frac{(\alpha+\beta)m_c^2}{\alpha\beta M_B^2}} \, .
\end{eqnarray}
For $\eta_9$:
\begin{eqnarray}
\nonumber
\rho^{pert}_9(s)&=&\frac{1}{2^8\pi^6}\int^{\alpha_{max}}_{\alpha_{min}}\frac{d\alpha}{\alpha^3}\int^{\beta_{max}}_{\beta_{min}}
\frac{d\beta}{\beta^3}(1-\alpha-\beta)^2[(\alpha+\beta)m_c^2-3\alpha\beta
s][(\alpha+\beta)m_c^2-\alpha\beta s]^3 \, ,
\\ \nonumber
\rho^{\langle\bar qq\rangle}_9(s)&=&0 \, ,
\\ \nonumber
\rho^{\langle GG\rangle}_9(s)&=&\frac{\langle
g_s^2GG\rangle}{2^8\pi^6}\int^{\alpha_{max}}_{\alpha_{min}}\frac{d\alpha}{\alpha^2}\int^{\beta_{max}}_{\beta_{min}}
d\beta\{
\frac{(1-\alpha-\beta)^2m_c^2}{3\alpha}[2(\alpha+\beta)m_c^2-3\alpha\beta
s]
\\ \nonumber
&&\hspace{4cm}+\frac{1-\alpha-\beta}{\beta}[(\alpha+\beta)m_c^2-2\alpha\beta s][(\alpha+\beta)m_c^2-\alpha\beta s]\} \, ,
\\ \nonumber
\rho^{mix}_9(s)&=&0 \, ,
\\
\rho^{\langle\bar qq\rangle^2}_9(s)&=&-\frac{m_c^2\langle\bar qq\rangle^2}{6\pi^2}\sqrt{1-4m_c^2/s} \, ,
\end{eqnarray}
\begin{eqnarray}
\nonumber
\Pi^{mix\langle\bar qq\rangle}_9(M_B^2)&=&-\frac{m_c^2\langle\bar qg_s\sigma\cdot Gq\rangle\langle\bar qq\rangle}{12\pi^2}\int_0^1
\frac{d\alpha}{\alpha}(\frac{m_c^2}{\alpha M_B^2}-1)e^{-\frac{m_c^2}{\alpha(1-\alpha)M_B^2}} \, ,
\\ \nonumber
\Pi^{\langle GGG\rangle}_9(M_B^2)&=&-\frac{\langle g_s^3fGGG\rangle}{3\times2^9\pi^6}\int_0^1d\alpha\int_0^{\beta_{max}}d\beta
\\ \nonumber
&& \{\frac{\ln(\alpha\beta(1-\alpha-\beta)M_B^4)-2\ln2-\gamma_E}{\alpha\beta}[12(\alpha\beta M_B^2)^2
+6\alpha\beta M_B^2(\alpha+\beta)m_c^2
+(\alpha+\beta)^2m_c^4]
\\ \nonumber
&&+\frac{33(\alpha\beta M_B^2)^2+12\alpha\beta M_B^2(\alpha+\beta)m_c^2+(\alpha+\beta)^2m_c^4}{\alpha\beta}
+\frac{(1-\alpha-\beta)^2m_c^2}{\alpha^4}[2\alpha\beta M_B^2+(\alpha+\beta)m_c^2]
\\ \nonumber
&&-\frac{(1-\alpha-\beta)^2}{2\alpha^3}[3\alpha\beta M_B^4+M_B^2(\alpha+\beta)m_c^2]\}e^{-\frac{(\alpha+\beta)m_c^2}{\alpha\beta M_B^2}} \, .
\end{eqnarray}
For $\eta_{10}$:
\begin{eqnarray}
\nonumber
\rho^{pert}_{10}(s)&=&\frac{3}{2^5\pi^6}\int^{\alpha_{max}}_{\alpha_{min}}\frac{d\alpha}{\alpha^3}\int^{\beta_{max}}_{\beta_{min}}
\frac{d\beta}{\beta^3}(1-\alpha-\beta)^2[(\alpha+\beta)m_c^2-3\alpha\beta
s][(\alpha+\beta)m_c^2-\alpha\beta s]^3 \, ,
\\ \nonumber
\rho^{\langle\bar qq\rangle}_{10}(s)&=&0 \, ,
\\ \nonumber
\rho^{\langle GG\rangle}_{10}(s)&=&\frac{\langle
g_s^2GG\rangle}{2^6\pi^6}\int^{\alpha_{max}}_{\alpha_{min}}\frac{d\alpha}{\alpha^2}\int^{\beta_{max}}_{\beta_{min}}
d\beta\{
\frac{2(1-\alpha-\beta)^2m_c^2}{\alpha}[2(\alpha+\beta)m_c^2-3\alpha\beta
s]
\\ \nonumber
&&\hspace{4cm}+\frac{5(1-\alpha-\beta)^2+10\alpha\beta}{4\beta^2}[(\alpha+\beta)m_c^2-2\alpha\beta s][(\alpha+\beta)m_c^2-\alpha\beta s]\\ \nonumber
&&\hspace{4cm}+\frac{6(1-\alpha-\beta)}{\beta}[(\alpha+\beta)m_c^2-2\alpha\beta s][(\alpha+\beta)m_c^2-\alpha\beta s]\} \, ,
\\ \nonumber
\rho^{mix}_{10}(s)&=&0 \, ,
\\
\rho^{\langle\bar qq\rangle^2}_{10}(s)&=&-\frac{4m_c^2\langle\bar qq\rangle^2}{\pi^2}\sqrt{1-4m_c^2/s} \, ,
\end{eqnarray}
\begin{eqnarray}
\nonumber
\Pi^{mix\langle\bar qq\rangle}_{10}(M_B^2)&=&-\frac{2m_c^2\langle\bar qg_s\sigma\cdot Gq\rangle\langle\bar qq\rangle}{\pi^2}\int_0^1
\frac{d\alpha}{\alpha}(\frac{m_c^2}{\alpha M_B^2}-1)e^{-\frac{m_c^2}{\alpha(1-\alpha)M_B^2}} \, ,
\\ \nonumber
\Pi^{\langle GGG\rangle}_{10}(M_B^2)&=&-\frac{\langle g_s^3fGGG\rangle}{3\times2^6\pi^6}\int_0^1d\alpha\int_0^{\beta_{max}}d\beta
\\ \nonumber
&&\{\frac{\ln(\alpha\beta(1-\alpha-\beta)M_B^4)-2\ln2-\gamma_E}{\alpha\beta}[12(\alpha\beta M_B^2)^2
+6\alpha\beta M_B^2(\alpha+\beta)m_c^2+(\alpha+\beta)^2m_c^4]
\\ \nonumber
&&+\frac{33(\alpha\beta M_B^2)^2+12\alpha\beta M_B^2(\alpha+\beta)m_c^2+(\alpha+\beta)^2m_c^4}{\alpha\beta}
+\frac{3(1-\alpha-\beta)^2m_c^2}{\alpha^4}[2\alpha\beta M_B^2+(\alpha+\beta)m_c^2]
\\ \nonumber
&&-\frac{3(1-\alpha-\beta)^2}{2\alpha^3}[3\alpha\beta M_B^4+M_B^2(\alpha+\beta)m_c^2]
+\frac{2}{1-\alpha-\beta}[3\alpha\beta M_B^4+M_B^2(\alpha+\beta)m_c^2]\}e^{-\frac{(\alpha+\beta)m_c^2}{\alpha\beta M_B^2}} \, .
\end{eqnarray}

\end{enumerate}


\begin{thebibliography}{100}

\bibitem{2003-Choi-p262001-262001}
  S.~K.~Choi et al.,
     Phys.\ Rev.\  Lett {\bf 91}, 262001 (2003).

\bibitem{2005-Choi-p182002-182002}
 S.~K.~Choi et al.,
     Phys.\ Rev.\  Lett {\bf 94}, 182002 (2005).


\bibitem{2005-Aubert-p142001-142001}
 B.~Aubert et al., BABAR Collaboration,
     Phys.\ Rev.\  Lett {\bf 95}, 142001 (2005).


\bibitem{2007-Yuan-p182004-182004}
 C.~Z.~Yuan et al., Belle Collaboration,
     Phys.\ Rev.\  Lett {\bf 99}, 182004 (2007).


\bibitem{2006-Uehara-p82003-82003}
 S.~Uehara et al.,
     Phys.\ Rev.\  Lett {\bf 96}, 082003 (2006).



\bibitem{2007-Abe-p82001-82001}
 K.~Abe et al.,
     Phys.\ Rev.\  Lett {\bf 98}, 082001 (2007).


\bibitem{2007-Aubert-p212001-212001}
 B.~Aubert et al., BABAR Collaboration,
     Phys.\ Rev.\  Lett {\bf 98}, 212001 (2007).


\bibitem{2007-Wang-p142002-142002}
 X.~L.~Wang et al., Belle Collaboration,
     Phys.\ Rev.\  Lett {\bf 99}, 142002 (2007).


\bibitem{2008-Choi-p142001-142001}
 S.~K.~Choi et al., Belle Collaboration,
     Phys.\ Rev.\  Lett {\bf 100}, 142001 (2008).


\bibitem{2008-Mizuk-p72004-72004}
 R.~Mizuk et al., Belle Collaboration,
     Phys.\ Rev.\  D {\bf 78}, 072004 (2008).

\bibitem{2009-Aaltonen-p242002-242002}
 T.~Aaltonen et al., CDF Collaboration,
     Phys.\ Rev.\  Lett {\bf 102}, 242002 (2009).


\bibitem{2006-Swanson-p243-305}
 E.~S.~Swanson,
     Phys.\ Rep. {\bf 429}, 243 (2006).


\bibitem{2008-Zhu-p283-322}
 S.~L.~Zhu,  Int. J. Mod. Phys. E {\bf 17}, 283 (2008).


\bibitem{2008-Amsler-p1}
 C.~Amsler et al., (Particle Data Group),
     Phys.\ Lett. B {\bf 667}, 1 (2008).

\bibitem{2009-Bracko-p-}
 M.~Bracko,
     arXiv:0907.1358 [hep-ex].

\bibitem{2003-Close-p210-216}
 F.~E.~Close and S.~Godfrey, Phys. Let. B {\bf{574}}, 210 (2003).


\bibitem{2004-Swanson-p197-202}
 E.~S.~Swanson, Phys.\ Lett. B {\bf 598}, 197 (2004);~ E.~S.~Swanson, Phys.\ Lett. B {\bf 588}, 189 (2004);~T.~Fernandez-Carames, A.~Valcarce, and J.~Vijande,  Phys.\ Rev.\  Lett {\bf 103}, 222001 (2009).

\bibitem{2007-Matheus-p14005-14005}
  R.~D.~Matheus, S.~Narison, M.~Nielsen, and J.~M.~Richard,
  Phys.\ Rev.\  D {\bf 75}, 014005 (2007);L.~Maiani,~A.~D.~Polosa, and V.~Riquer,   Phys.\ Rev.\  Lett {\bf 99}, 182003 (2007).


\bibitem{2008-Liu-p94015-94015}
 X.~Liu,~Y.R.~Liu,~W.~Z~Deng, and S.~L. Zhu,  Phys.\ Rev.\  D {\bf 77}, 094015 (2008);
 ~ X.~Liu,~Y.R.~Liu,~W.~Z~Deng, and S.~L. Zhu,  Phys.\ Rev.\  D {\bf 77}, 034003 (2008);
 ~Su Houng Lee, ~A.~Mihara,~F.~ Navarra, and M.~ Nielsen,  Phys. Lett. B {\bf{661}}, 28 (2008);
 ~C.~Meng and K.~T.~Chao,~arXiv:0708.4222[hep-ph];~G.~J.~Ding,~arXiv:0711.1485[hep-ph].

\bibitem{2009-Bracco-p240-244}
  M.~E.~Bracco, S.~H.~Lee, M.~Nielsen, and R.~Rodrigues da Silva,  Phys. Lett. B {\bf{671}}, 240 (2009);~L.~Maiani,~A.~D.~Polosa, and V.~Riquer,  New J. Phys. {\bf{10}}(2008) 073004.


\bibitem{2005-Close-p215-222}
 F.~E.~Close and P.~R.~Page,  Phys. Lett. B {\bf{628}}, 215 (2005).


\bibitem{2005-Maiani-p031502-031502}
 L.~Maiani, V.~Riquer, F.~Piccinini, and A.~D.~Polosa,  Phys.\ Rev.\  D {\bf 72}, 031502 (2005).


\bibitem{2009-Liu-p17502-17502}
 X.~Liu and S.~L.Zhu,  Phys.\ Rev.\  D {\bf 80}, 017502 (2009);~N.~Mahajan  Phys.\ Lett.\  B {\bf 679}, 228 (2009);~T.~Branz,~T.~Gutsche, and V.~E.~Lyubovitskij, Phys.\ Rev.\  D {\bf 80}, 054019 (2009);~G.~J.~Ding, Eur.\ Phys.\ J.\ C {\bf 64} 297 (2009).


\bibitem{libaiqing}
Bai-Qing Li and Kuang-Ta Chao, Phys. Rev. D {\bf 79}, 094004 (2009).


\bibitem{1980-Chao-p281-281}
    Kuang-Ta Chao, Nucl. \ Phys. \  B {\bf{169}}, 281(1980);~Nucl. \ Phys. \  B {\bf 183}, 435(1981).


\bibitem{2009-Jiao-p114034-114034}
  Chun-Kun Jiao, Wei Chen, Hua-Xing Chen, and Shi-Lin Zhu,
    Phys.\ Rev.\  D {\bf 79}, 114034 (2009).



\bibitem{M.A.Shifman:1979wz}
  M.~A.~Shifman, A.~I.~Vainshtein, and V.~I.~Zakharov,
  Nucl.\ Phys.\ B {\bf 147}, 385 (1979).

\bibitem{L.J.Reinder:1985wz}
  L.~J.~Reinders, H.~Rubinstein, and S.~Yazaki,
  Phys.\ Rept.\  {\bf 127}, 1 (1985).

\bibitem{Colangelo:2000dp}
  P.~Colangelo and A.~Khodjamirian,
  in \textit{At the Frontier of Particle Physics/Handbook of QCD}, M.~Shifman (World
  Scientific, Singapore, 2001) Vol. {\bf 3}, p. 1495.


\bibitem{Matheus:2009vq}
  R.~D.~Matheus, F.~S.~Navarra, M.~Nielsen, and C.~M.~Zanetti,
  Phys.\ Rev.\  D {\bf 80}, 056002 (2009).

\bibitem{Zhang:2009vs}
  J.~R.~Zhang and M.~Q.~Huang,
   Phys.\ Rev.\  D {\bf 80}, 056004 (2009).

\bibitem{2001-Eidemuller-p203-210}
   M.~Eidermuller and M.~Jamin,
   Phys.\ Lett.\  B {\bf{498}}, 203 (2001).

\bibitem{1999-Jamin-p300-303}
   M.~Jamin and A.~Pich,
   Nucl.\ Phys.\ Proc.\ Suppl. {\bf{74}}, 300 (1999).

\bibitem{2009-Lee-p29-39}
   S.~H.~Lee, K.~Morita, and M.~Nielsen,
  Nucl. Phys, A {\bf{815}}, 29 (2009).

\bibitem{2009-Wang-p375-382}
 Zhi-Gang Wang, Eur. Phys. J, C {\bf 62}, 375 (2009);~Phys.\ Rev.\  D {\bf 79}, 094027 (2009).


\end{thebibliography}
\end{document}